\documentclass[preprint]{ptephy_v2}

\usepackage{mathtools}
\usepackage{siunitx}
\usepackage[hidelinks]{hyperref}

\makeatletter
\renewcommand{\collaborator}[1]{%
  \g@addto@macro\AB@authlist{%
    \protect\\[0.8em]\protect\normalfont\protect\textbf{#1}}%
}
\makeatother

\IfFileExists{ptephy.bst}{\bibliographystyle{ptephy}}{\bibliographystyle{unsrt}}

\newcommand{\meV}{\,\mathrm{meV}}

\newcommand{\Hz}{\,\mathrm{Hz}}
\newcommand{\kHz}{\,\mathrm{kHz}}
\newcommand{\MHz}{\,\mathrm{MHz}}
\newcommand{\GHz}{\,\mathrm{GHz}}

\newcommand{\second}{\,\text{s}}
\newcommand{\minutes}{\,\text{min}}

\newcommand{\dB}{\,\text{dB}}

\newcommand{\W}{\,\text{W}}

\newcommand{\mm}{\,\text{mm}}

\newcommand{\sqmm}{\,\text{mm}^2}

\newcommand{\K}{\,\text{K}}
\newcommand{\kmpers}{\,\text{km}/\text{s}}
\newcommand{\Deg}{^{\,\circ}}
\newcommand{\GeVpercubcm}{\,\mathrm{GeV}/\text{cm}^3}

\newcommand{\mDP}{m_\text{DP}}
\newcommand{\dnu}{\varDelta\nu}
\newcommand{\kB}{k_\text{B}}

\newcommand{\vE}{v_\text{E}}
\newcommand{\vc}{v_\text{c}}
\newcommand{\vDP}{v_\text{DP}}

\newcommand{\PDP}{P_\text{DP}}
\newcommand{\Pout}{P_\text{out}}
\newcommand{\Pin}{P_\text{in}}
\newcommand{\Trx}{T_\text{rx}}
\newcommand{\Aeff}{A_\text{eff}}

\newcommand{\nuRF}{\nu_\text{RF}}
\newcommand{\nuLO}{\nu_\text{LO}}
\newcommand{\nuIF}{\nu_\text{IF}}
\newcommand{\nuUSB}{\nu_\text{USB}}

\newcommand{\searchband}{$242.94935$--$242.94983\GHz$}
\newcommand{\massband}{$1.0047578$--$1.0047598\meV$}

\newcommand{\yband}{$170$--$260\GHz$}
\newcommand{\ymassband}{$0.7$--$1.1\meV$}

\newcommand{\chilimit}{(1.6\text{--}1.9)\times10^{-9}}

\begin{document}

\title{Search for dark photon dark matter near $\mathbf{1\,meV}$ using a superconducting mixer}

\author[1,*]{K.~Hanimura}
\affil{Department of Physics, Graduate School of Science, Kyoto University,
Kyoto 606-8502, Japan \email{hanimura.keigo.58c@st.kyoto-u.ac.jp}}

\author[2,3]{S.~Adachi}
\affil[2]{Department of Physics, Okayama University, Okayama 700-8530, Japan}
\affil[3]{High Energy Accelerator Research Organization (KEK), Ibaraki 305-0801, Japan}

\author[4]{Y.~Hasegawa}
\affil[4]{National Institute of Information and Communications Technology,
Tokyo 184-8795, Japan}

\author[5]{S.~Honda}
\affil[5]{Division of Physics, Faculty of Pure and Applied Sciences,
University of Tsukuba, Ibaraki 305-8571, Japan}

\author[6]{T.~Nakajima} 
\affil[6]{Department of Mechanical and Electrical Engineering, Faculty of Engineering, Suwa University of Science, Nagano 391-0292, Japan}

\author[7]{H.~Ogawa}
\affil[7]{Graduate School of Science, Osaka Metropolitan University, Osaka 558-8585, Japan}

\author[1]{J.~Suzuki}
\author[1]{O.~Tajima}
\author[1]{H.~Takeuchi}

\collaborator{The DOSUE-RR Collaboration}

\begin{abstract}%
We demonstrate the use of a superconducting mixer for direct searches for dark photon dark matter (DPDM) in the $\meV$ mass range.
DPDM couples to the ordinary electromagnetic field through kinetic mixing with a coupling constant $\chi$ and can convert into ordinary photons at the surface of a metal plate.
For DPDM masses of $1\meV$, the conversion photons have frequencies of around $240\GHz$.
At such frequencies, superconducting mixers enable low-noise spectroscopic measurements.
We performed a direct search for DPDM in the mass range \massband, corresponding to the frequency range \searchband, using approximately one hour of data obtained with a cryogenic millimeter-wave receiver based on a superconducting mixer.
No significant excess was found, and we set upper limits of $\chi < \chilimit$ at the 95\% confidence level.
These limits are comparable to existing constraints from solar dark photons.
By applying this method, we plan to extend the search to the wider mass range of \ymassband, corresponding to \yband, with improved sensitivity.

\end{abstract}

\subjectindex{F40, F41, H20}

\maketitle

\textit{1. Introduction.  }
Determining the nature of dark matter remains a central open question in particle physics and cosmology.
Dark photon dark matter (DPDM) is a dark matter candidate described by a vector field with mass $\mDP$.
This field couples to the ordinary electromagnetic field through kinetic mixing with a coupling constant $\chi$~\cite{Wispy}.
Searches for DPDM in the $\mu\mathrm{eV}$--$\mathrm{meV}$ mass range are motivated by high-scale inflationary production scenarios and by certain string-theory models~\cite{inflation,Wispy}.

Through kinetic mixing, DPDM can convert into ordinary photons at an interface between media~\cite{AxionLimit}.
This conversion effect enables searches in which photons emitted from a metal surface are received by an antenna~\cite{DishAntenna}.
The conversion photons propagate nearly along the surface normal, deviating from it by only $\lesssim 0.1\Deg$, because DPDM in the Galactic halo is nonrelativistic, with $\vDP/c\approx10^{-3}$~\cite{DishDirection}.
A conversion-photon signal would appear as a narrow peak in the power spectrum.
The peak frequency $\nu$ is determined primarily by the DPDM rest-mass energy, $h\nu\simeq\mDP c^2$, while the fractional linewidth, of order $10^{-6}$, is determined by the Galactic velocity distribution~\cite{PhysRevLett.51.1415}.

Following Ref.~\cite{DishAntenna}, the expected conversion-photon power received by the antenna, $\PDP$, is given by
\begin{eqnarray}
    \PDP = 1.1 \times 10^{-19}\W
    \left(\frac{\chi}{10^{-9}}\right)^{2}
    \left(\frac{\Aeff}{18\sqmm}\right)
    \left(\frac{\rho}{0.39\GeVpercubcm}\right)
    \left(\frac{\alpha}{\sqrt{1/3}}\right)^2 .
    \label{eq:power}
\end{eqnarray}
%
Here, $\Aeff$ is the effective aperture area of the antenna,
$\rho=0.39\GeVpercubcm$ is the local dark matter density in the Galactic halo~\cite{DMdensity}, and $\alpha=\sqrt{1/3}$ is the polarization factor for a single-polarization receiver, assuming randomly oriented DPDM polarization.


In the $\meV$ mass range, indirect constraints and recent direct searches have been reported~\cite{solar,Wispy,PlaneParabolic,Dandelion}.
In this work, we present a direct search for DPDM near $1\meV$, at a frequency of approximately $243\GHz$.
This measurement demonstrates that a low-noise direct search at the $\meV$ mass scale is feasible using a cryogenic receiver based on a superconductor--insulator--superconductor (SIS) mixer.

At millimeter-wave frequencies below about $100\GHz$, low-noise spectroscopy can be performed using cryogenic semiconductor amplifiers~\cite{DOSUE-K}.
However, above about $100\GHz$, low-noise spectroscopy based on semiconductor amplification becomes challenging.
%
For example, a commercially available amplifier covering $200$--$260\GHz$ has a noise figure exceeding $6\dB$, corresponding to a noise temperature above approximately $860\K$~\cite{VDI_LNA}.
This challenge can be addressed by placing an SIS mixer upstream of the first amplifier in the receiver chain.
SIS mixers incorporate SIS junctions, each consisting of two superconducting electrodes separated by a thin insulator.
Quasiparticle tunneling through the insulator produces a strongly nonlinear current--voltage characteristic, which enables low-noise frequency conversion~\cite{TuckerFeldman}.
State-of-the-art SIS mixers achieve noise temperatures within several times the quantum limit, $h\nu/\kB$ (approximately $11.7\K$ at $243\GHz$)~\cite{SIS2007}.
In the mixer, the radio-frequency (RF) signal is combined with a local oscillator (LO) signal  and down-converted to an intermediate frequency (IF), $\nuIF=\left|\nuRF-\nuLO\right|$.
In this way, a millimeter-wave signal above $100\GHz$ is converted to an IF signal around $10\GHz$, where cryogenic low-noise amplifiers with noise temperatures of approximately $4\K$ are available.
This method enables low-noise spectroscopy of the original RF signal.

\begin{figure}[t]
\centering
\includegraphics[width=\linewidth]{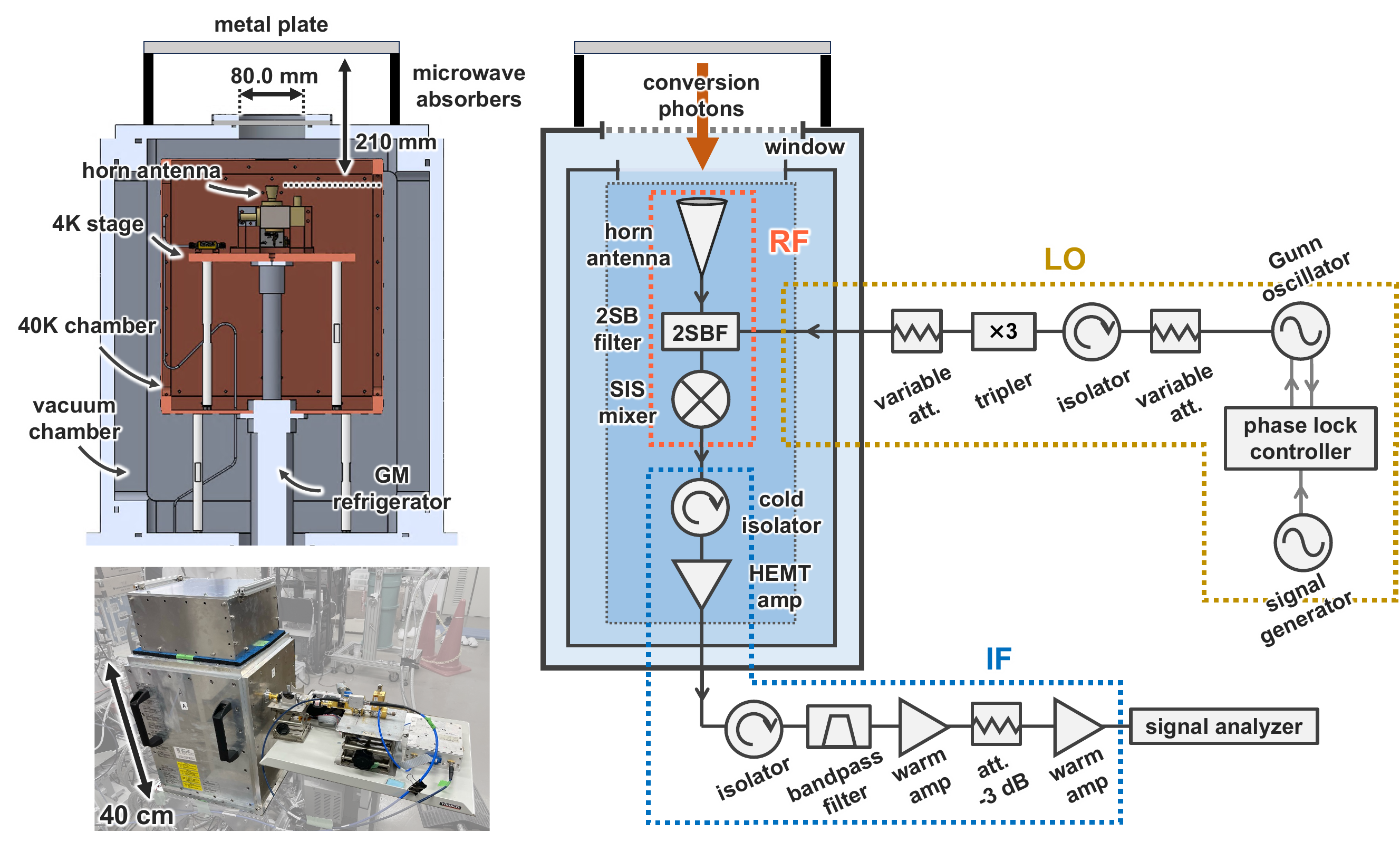}
\caption{\label{fig:CAD_photo_diagram}
Photograph and schematic overview of the experimental setup.
Dark photons convert into ordinary photons at the surface of an aluminum plate.
The conversion photons are received by a horn antenna.
The RF signal passes through a sideband-separating (2SB) filter and is down-converted by an SIS mixer before amplification and spectral measurement.}
\end{figure}

\textit{2. Experimental setup.  }
To demonstrate this concept, we developed a cryogenic millimeter-wave receiver based on an SIS mixer, as shown in Fig.~\ref{fig:CAD_photo_diagram}.

\begin{figure}[t]
\centering
\begin{minipage}[c]{0.49\textwidth}
\centering
\includegraphics[width=\linewidth]{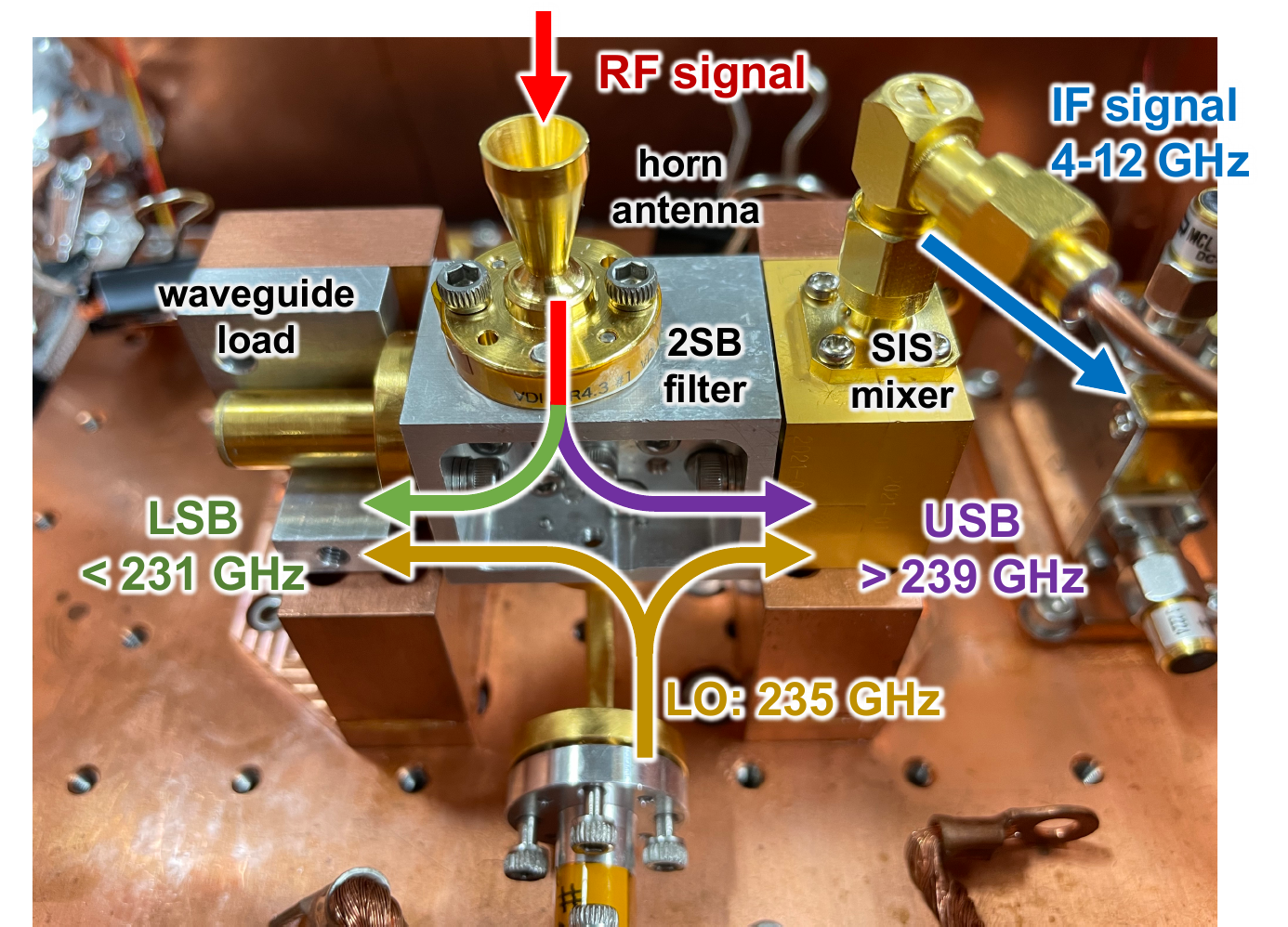}
\end{minipage}
\hfill
\begin{minipage}[c]{0.5\textwidth}
\centering
\includegraphics[width=\linewidth]{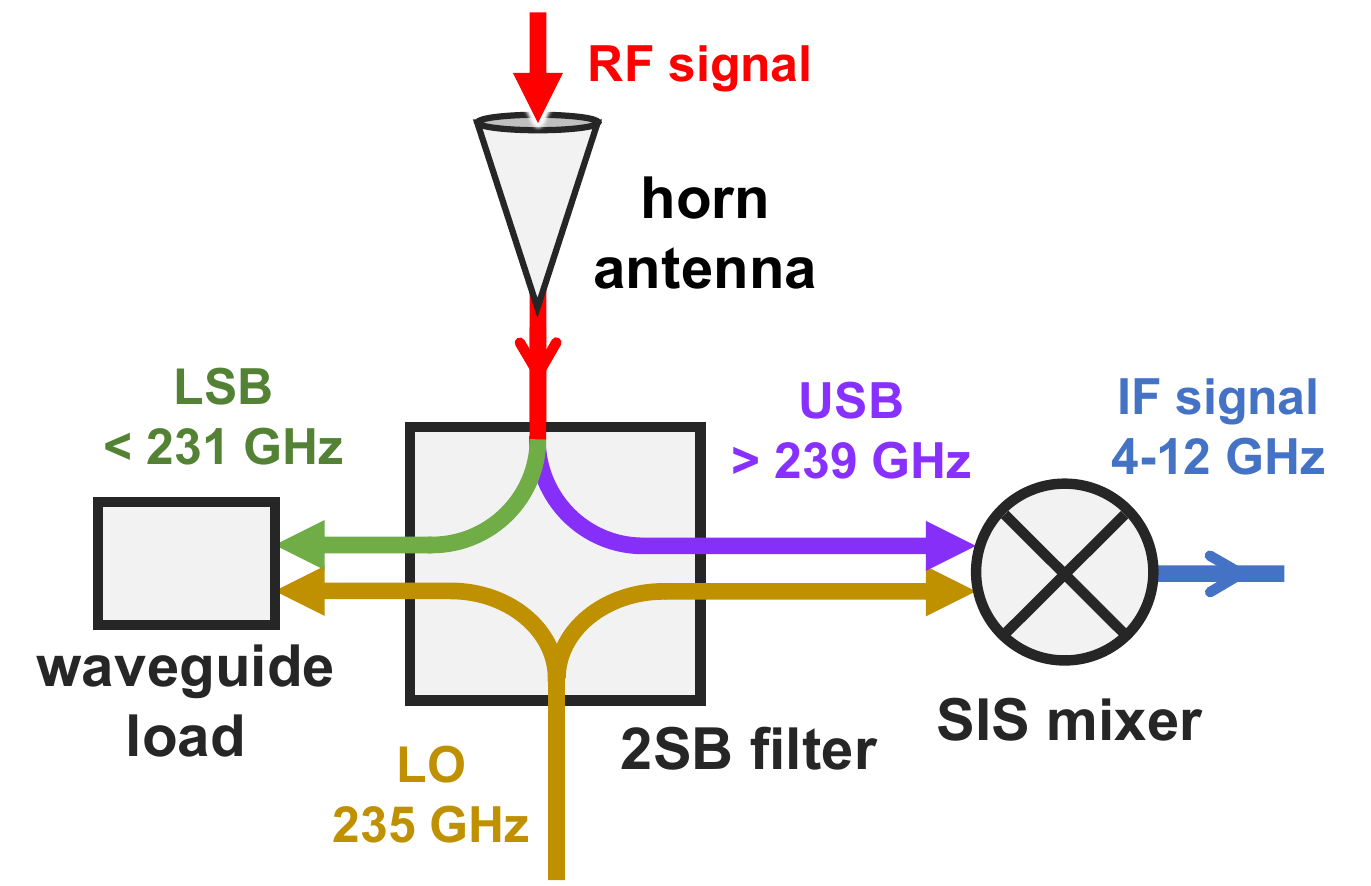}
\end{minipage}
\caption{\label{fig:rfchain}
RF section from the horn antenna to the SIS mixer.
The photograph shows the cryogenic RF components, and the block diagram shows the signal path through the 2SB filter.}
\end{figure}

\begin{figure}[t]
\centering
\includegraphics[width=0.5\linewidth]{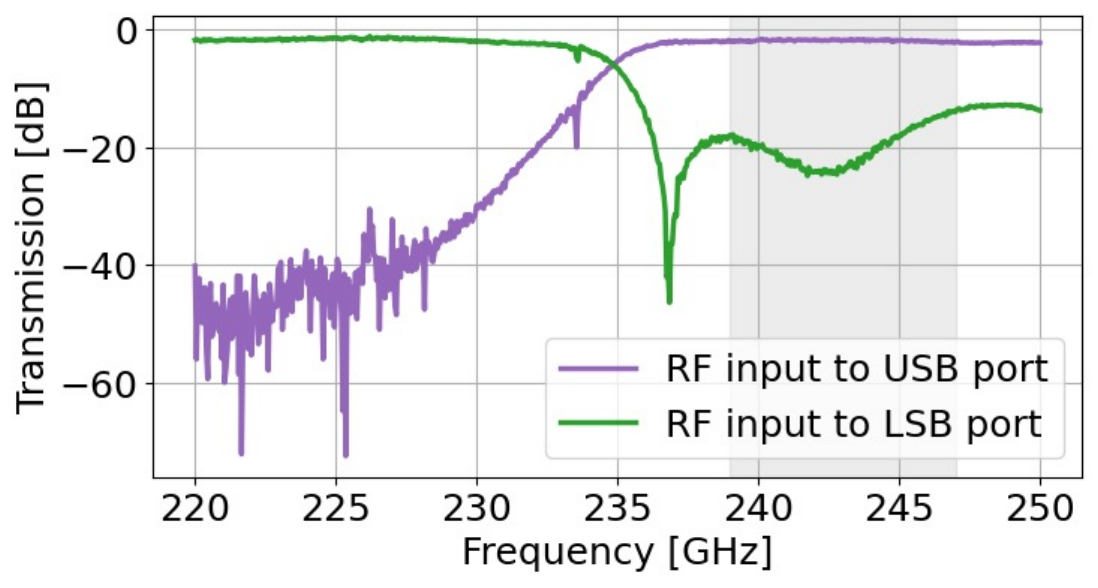}
\caption{\label{fig:2SB_S21}
Frequency response of the 2SB filter.
The purple and green curves show the transmission from the RF port to the USB and LSB output ports, respectively.
The gray shaded region indicates the accessible RF range of the receiver.
}
\end{figure}

The RF section from the horn antenna to the SIS mixer is shown in Fig.~\ref{fig:rfchain}.
We used a WR-4.3 conical horn antenna (VDI Inc.) with an aperture diameter of $7.1\mm$.
The RF signal received by the horn antenna was guided to a sideband-separating (2SB) filter~\cite{2SBF,Asayama2015IRF}.
The 2SB filter separated the input RF signal into the upper sideband (USB, $> 239\GHz$) and the lower sideband (LSB, $< 231\GHz$) with respect to the filter separation frequency ($235 \GHz$), as shown in Fig.~\ref{fig:2SB_S21}.
Compared with the conventional separation method using a 2SB SIS mixer, this scheme has been shown to provide a higher and more stable image rejection ratio (IRR), which quantifies the suppression of signals from the opposite sideband~\cite{2SBF,Asayama2015IRF}.
Over the receiver’s accessible RF range of $239$--$247\GHz$, the IRR exceeded $24.0 \dB$, corresponding to a leakage of LSB power into the USB output port of less than $0.40\%$.
The LO frequency was set to the separation frequency, $\nuLO=235\GHz$.
The LO signal was generated by tripling the approximately $78.33\GHz$ output of a Gunn oscillator (Nitsuki 3845E) using a tripler (VDI Inc. WR4.3X3S).
The LO was phase-locked with a phase-locked loop (PLL) unit (XL Microwave Inc. Model 800A).
The PLL used a reference signal from a signal generator (Keysight E8257D) with a fractional frequency accuracy better than $4\times10^{-8}$~\cite{KeysightE8257D}.
The corresponding frequency uncertainty (approximately $9 \kHz$) was much smaller than the expected signal linewidth.
To demonstrate this concept with a simple receiver configuration, only the USB was used.
The LSB output port was terminated with a waveguide load (Custom Microwave Inc. LT4R).
The USB signal and the LO signal were then mixed in the SIS mixer (T. Nakajima et al., in preparation), which produced the IF signal at the difference frequency.
This SIS mixer contains four SIS junctions and covers a broad RF range of \yband.

The IF signal then passed through a cryogenic isolator (Quinstar QCI-G8401201AMZBC) and a HEMT amplifier (Low Noise Factory LNF-LNC6\_20C).
The isolator suppressed standing waves between the SIS mixer and the HEMT amplifier.
The antenna, 2SB filter, SIS mixer, cryogenic isolator, and HEMT amplifier were mounted on a copper stage cooled to $4\K$ inside a chamber cooled to $40\K$.
The subsequent room-temperature chain consisted of an isolator (MTC 1166L2FF-S), a $4$--$12\GHz$ bandpass filter (Sogo Electronics Inc. SCBPF-8000/8000-O/O-A1), and two warm amplifiers (Mini-Circuits ZX60-02203+).
The receiver IF band was $4$--$12\GHz$.
With $\nuLO=235\GHz$, this IF band corresponds to an accessible RF range of $239$--$247\GHz$ because $\nuUSB=\nuLO+\nuIF$.
The IF signal was finally measured with a signal analyzer (Anritsu MS2840A).

An aluminum plate was mounted above the cryostat window to serve as the conversion surface.
The plate measured $280\mm\times280\mm$ and was $5\mm$ thick.
The window had a diameter of $80.0\mm$ and was sealed with a 25-$\mu\text{m}$-thick Kapton film.
The path of the conversion photons between the plate and the window was surrounded by microwave absorbers (Eccosorb AN-73).
Conversion photons produced at the plate surface passed through the window and were received by the horn antenna.
The distance between the plate and the antenna aperture was $210\mm$.
To suppress the radiative heat load, we installed radio-transparent multilayer insulation between the window and the antenna~\cite{RTMLI}.

\begin{figure}[t]
\centering
\begin{minipage}[c]{0.49\textwidth}
\centering
\includegraphics[width=\linewidth]{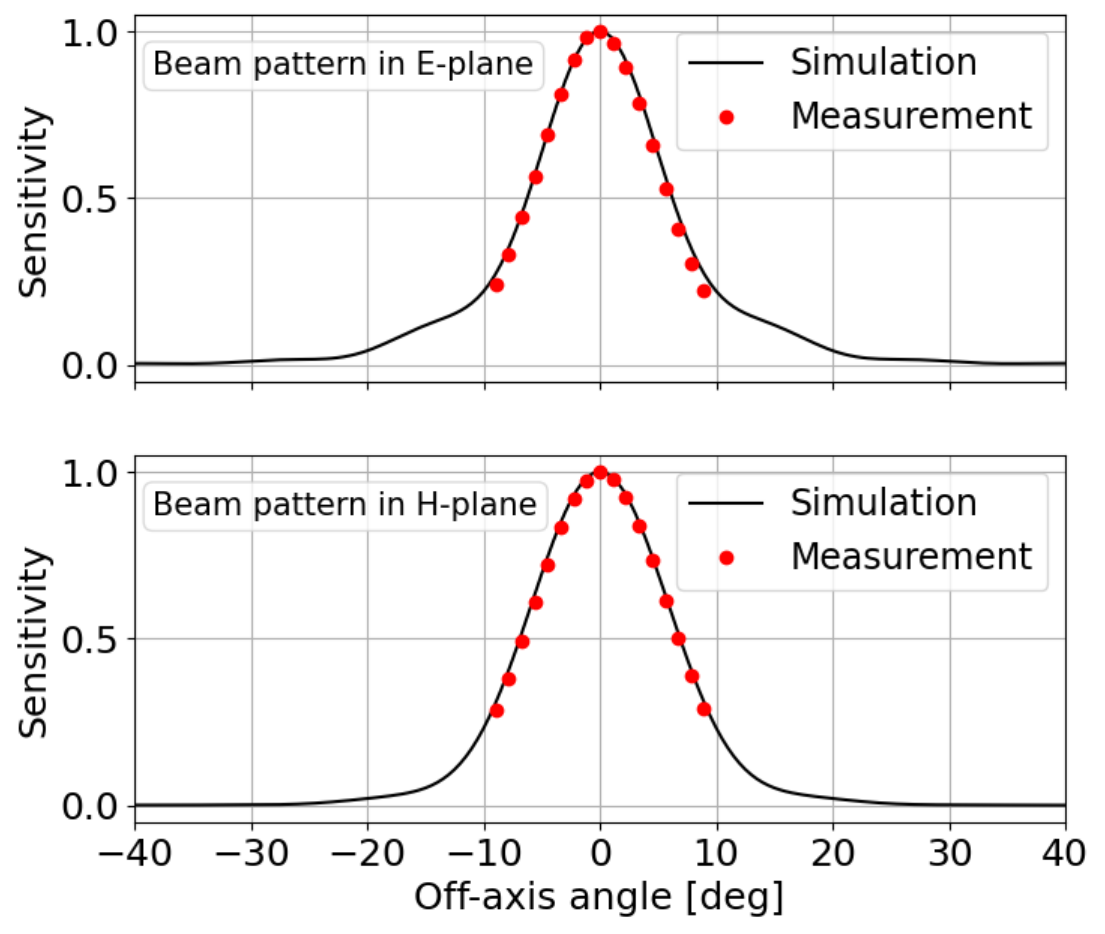}
\end{minipage}
\hfill
\begin{minipage}[c]{0.49\textwidth}
\centering
\includegraphics[width=\linewidth]{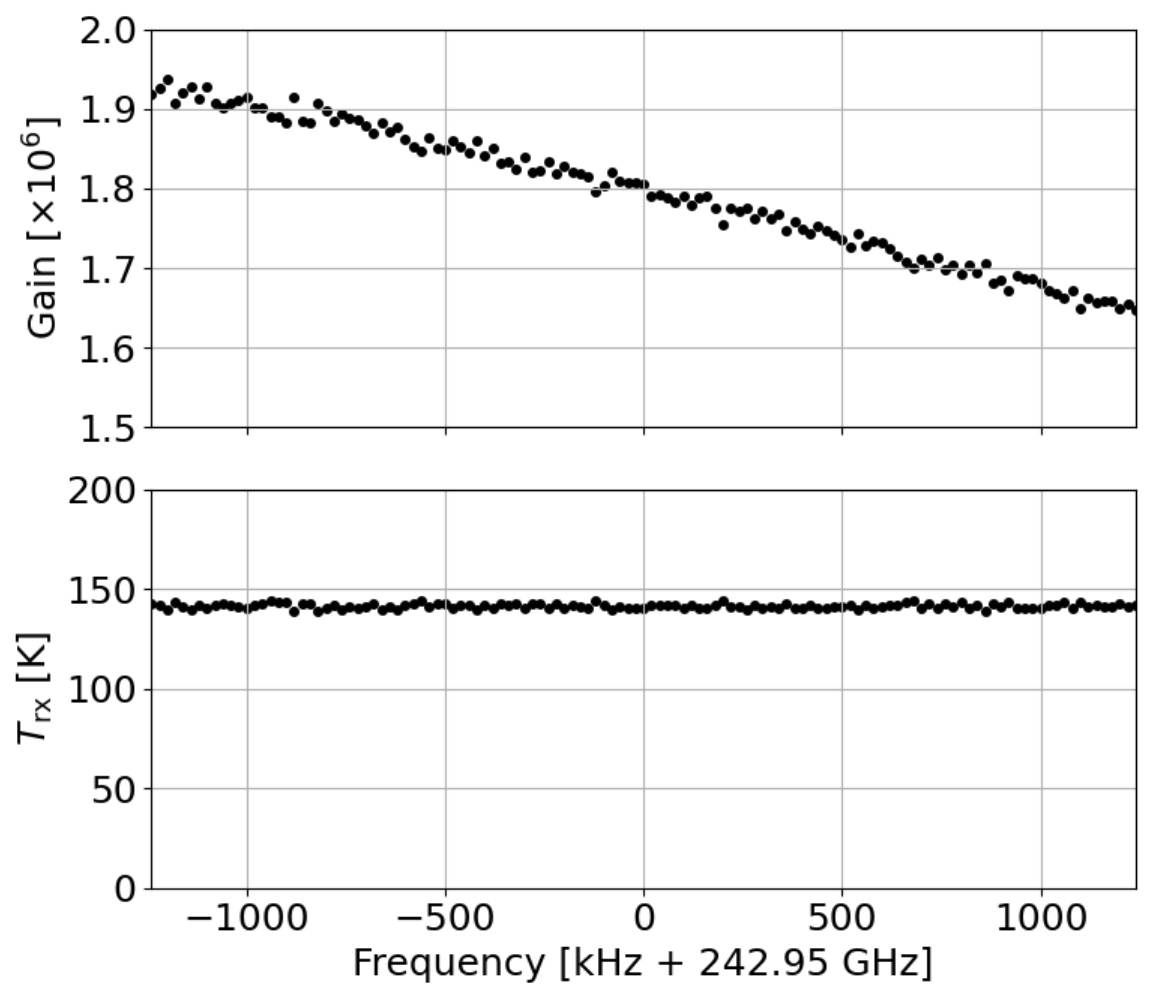}
\end{minipage}
\caption{\label{fig:bp_gain_Trx}
Left: simulated and measured beam patterns of the horn antenna in the E-plane (top) and H-plane (bottom).
Right: receiver gain $G$ (top) and noise temperature $\Trx$ (bottom) as functions of frequency.}
\end{figure}

\textit{3. Receiver characterization.  }
We characterized the receiver for the DPDM search by evaluating the effective aperture area of the antenna and by calibrating the receiver gain and noise temperature.
The effective aperture area at the horn mouth, $\Aeff$, was determined from its angular response (beam pattern).
We calculated the beam pattern with a three-dimensional electromagnetic simulation using Ansys HFSS, as shown in the left panel of Fig.~\ref{fig:bp_gain_Trx}~\cite{DOSUE-J}.
From the simulated beam pattern, we obtained $\Aeff = 18.4 \sqmm$ at $242.9493 \GHz$ (aperture efficiency 47\%) following the same procedure as in Ref.~\cite{DOSUE-J}.
To validate the simulation, we measured the beam pattern of the antenna at room temperature using a vector network analyzer (Keysight N5224B, VDI WR-4.3 extender) and an identical horn antenna~\cite{DOSUE-K}.
We compared the half-power beam widths (HPBWs) from the measured and simulated beam patterns.
The measured HPBWs were $12.3\Deg$ in the E-plane and $13.4\Deg$ in the H-plane; the corresponding simulated values were $12.8\Deg$ and $13.7\Deg$, respectively.
The difference in HPBW was 4.3\% for the E-plane and 2.2\% for the H-plane.
These differences were treated as the uncertainty in the beam width and propagated through $\Aeff$ to the systematic uncertainty in $\chi$.

The power measured by the signal analyzer, $\Pout$, is expressed as the receiver gain multiplied by the sum of the input power $\Pin$ and the receiver noise power:
\begin{equation}
 \Pout(\nu)=G(\nu)
 \left[\Pin(\nu)+\kB\Trx(\nu)\dnu\right],
 \label{eq:pout}
\end{equation}
where $G$ is the receiver gain, $\kB$ is the Boltzmann constant, $\Trx$ is the receiver noise temperature, and $\dnu$ is the bandwidth.
To obtain $G$ and $\Trx$, we performed a calibration using hot (room-temperature) and cold ($77\K$) blackbody radiation sources (Eccosrb CV-3).
The detailed calibration procedure followed Ref.~\cite{DOSUE-J}.
As shown in the right panel of Fig.~\ref{fig:bp_gain_Trx}, the receiver gain
and noise temperature in the selected search band were approximately $60\dB$ and $140\K$, respectively.
This noise temperature corresponds to approximately twelve times the quantum limit.
It is substantially lower than the noise temperatures of the commercial amplifiers, such as that in Ref.~\cite{VDI_LNA}.

\textit{4. Data taking and analysis.  }
We performed a DPDM search for one hour on December 6, 2024, at Kyoto University ($35.03^\circ$N, $135.78^\circ$E).
The polarization axis of the antenna was oriented at an azimuth of $97^\circ$, measured eastward from north.
We acquired 1800 spectra, each with an accumulation time of $2\second$.
For the analysis, we used 1580 spectra, corresponding to approximately $53\minutes$ of data, because the output power remained stable during this period.
Each spectrum covered $242.94875$--$242.95125\GHz$, a span of $2500\kHz$, which is the maximum span available in the FFT mode of the signal analyzer.
The resolution bandwidth was $300\Hz$.
Each spectrum consisted of 32,769 frequency points, with a frequency spacing of $76.3\Hz$.
Before and after the DPDM search, we calibrated the receiver gain and noise temperature by measuring the hot and cold loads for $10\second$ each.

\begin{figure}[t]
\centering
\begin{minipage}[c]{0.49\textwidth}
\centering
\includegraphics[width=\linewidth]{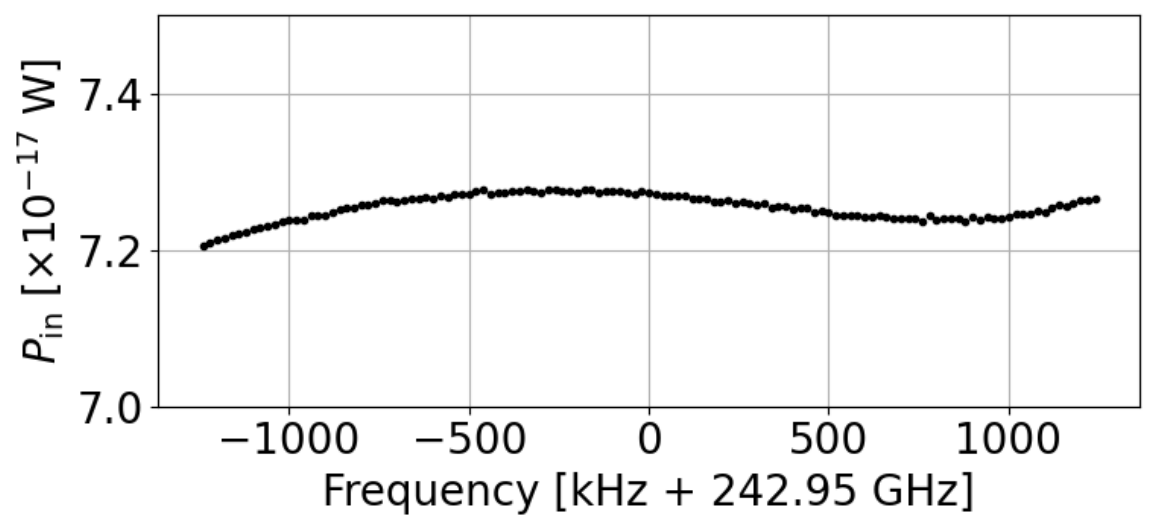}
\end{minipage}
\hfill
\begin{minipage}[c]{0.49\textwidth}
\centering
\includegraphics[width=\linewidth]{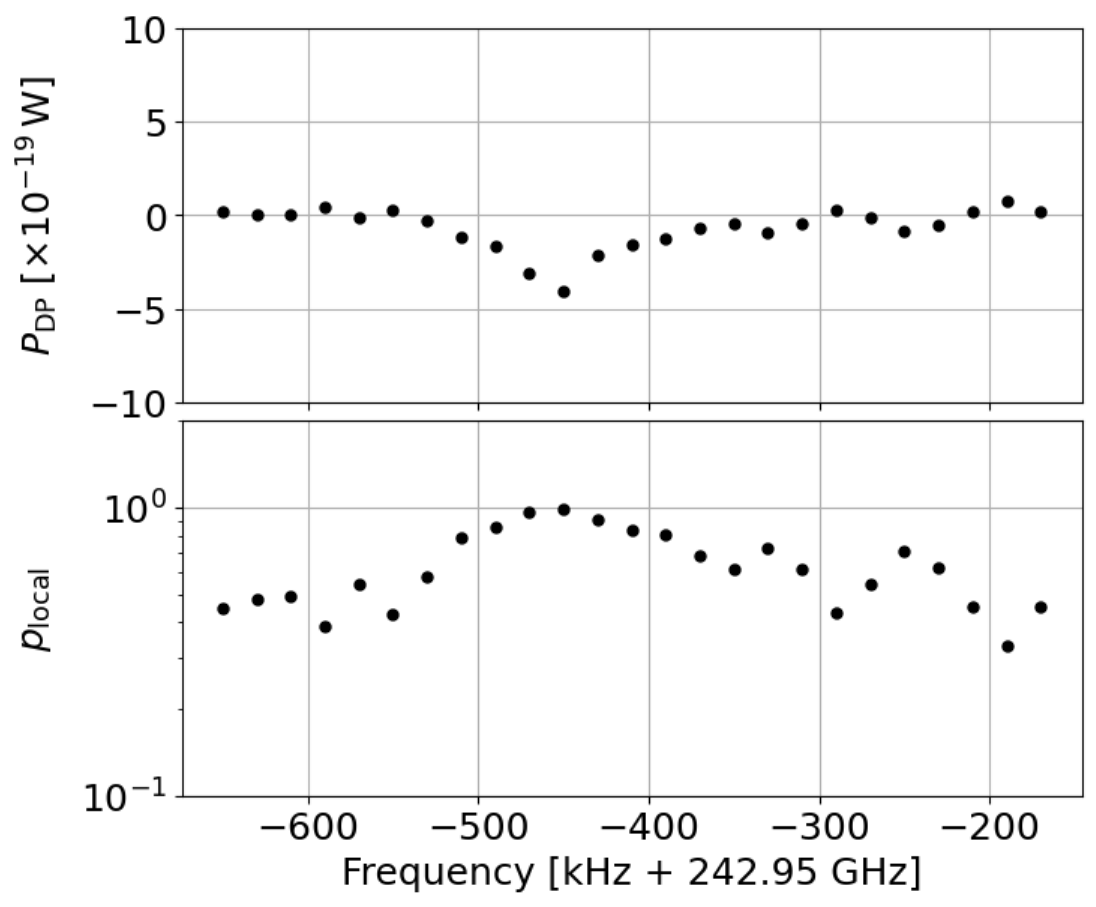}
\end{minipage}
\caption{\label{fig:Pin_Pdp_plocal}
Left: receiver input power spectrum $\Pin$ after time averaging and frequency rebinning to $\dnu=20\kHz$.
Right: extracted signal power and local $p$-value as functions of the assumed signal frequency $\nu_0$.
The local $p$-values were evaluated using null samples formed by subtracting pairs of spectra.}
\end{figure}

The analysis procedure follows that used in previous DOSUE-RR searches~\cite{DOSUE-J,DOSUE-K}, with modifications for the present data.
The procedure and analysis parameters specific to this measurement are summarized below.
The 1580 spectra were first averaged in time and then rebinned in frequency to a bin width of $\dnu=20\kHz$.
The rebinned spectrum was converted to the receiver input power $\Pin$ using Eq.~(\ref{eq:pout}).
%
In this conversion, the gain and receiver noise temperature were each modeled as a linear function of frequency over the full $2500\kHz$ span.
The resulting $\Pin$ spectrum is shown in the left panel of Fig.~\ref{fig:Pin_Pdp_plocal}.


To extract the signal power of the conversion photons, we fitted a function to the $\Pin$ spectrum at each assumed signal frequency $\nu_0 \equiv \mDP c^2/h$.
The fitting function at $\nu_0$ consists of a signal model and a baseline:
\begin{equation}
 f(\nu_i;\PDP,\{c_j\}) =
 \PDP f_{\rm sig}(\nu_i;\nu_0)
 +\sum_{j=0}^{3}c_j(\nu_i-\nu_0)^j .
 \label{eq:fit}
\end{equation}
Here, the first term represents the expected signal, and the second term is a polynomial describing the baseline.
The signal template was defined as the difference between the values of the cumulative distribution function at the two bin edges:
\begin{equation}
 f_{\rm sig}(\nu_i;\nu_0)=
 F(\nu_i+\dnu/2;\nu_0)-F(\nu_i-\dnu/2;\nu_0).
 \label{eq:template}
\end{equation}
\begin{align}
 F(\nu;\nu_0) &=
 \int_{0}^{v(\nu;\nu_0)} \mathrm{d}v'
 \int \mathrm{d}\Omega\,
 g(\mathbf{v}';\vc,\mathbf{v}_{\rm E})\,v'^2 ,
 \label{eq:fsig_cdf}\\
 g(\mathbf{v};\vc,\mathbf{v}_{\rm E}) &=
 \frac{1}{(\sqrt{\pi}\vc)^3}
 \exp\left[-\frac{|\mathbf{v}+\mathbf{v}_{\rm E}|^2}{\vc^2}\right].
 \label{eq:velocity_distribution}
\end{align}
Here $F(\nu;\nu_0)$ is the cumulative distribution function of the DPDM speed, evaluated at
$v(\nu;\nu_0) = c\sqrt{1-\left(\frac{\nu_0}{\nu}\right)^2}$.
The function $g$ is the DPDM velocity distribution,
$\vc$ is the rotational speed of the Galaxy, and $\mathbf{v}_{\rm E}$ is the velocity of the Earth in the Galactic frame.
We assumed a Maxwell--Boltzmann distribution with $|\mathbf{v}_{\rm E}|=\vc=220\kmpers$, as in many dark matter
searches~\cite{DMVelocityDistribution,DMVelocity}.

We scanned the assumed signal frequency $\nu_0$ over \searchband{} in steps of $20\kHz$.
At each step, the fit was performed over the range from $\nu_0-100\kHz$ to $\nu_0+900\kHz$, with $\PDP$ and $\{c_j\}$ as free parameters.
To estimate the uncertainty of each data point, we used the $500\kHz$ regions immediately below and above the fit range.
In each region, we subtracted a baseline obtained from a third-order polynomial fit and calculated the standard deviation of the residuals.
The average of these two standard deviations was assigned as the uncertainty of each data point in the fit.


The statistical significance was evaluated using null samples, following the procedure of Refs.~\cite{DOSUE-J,DOSUE-K}.
In the present analysis, null samples were prepared from the calibrated spectra by subtracting one $2\second$ spectrum from another.
Any stationary DPDM signal cancels out in the difference, while uncorrelated noise remains.
From the 1580 spectra, we formed $\binom{1580}{2}=1{,}247{,}410$ null samples.
The same template fit was applied to the null samples to obtain the null distribution of $x=\PDP/\sigma$, where $\sigma$ is the fit uncertainty in $\PDP$.
Because a DPDM signal can produce only a positive excess, the local $p$-value for each $\nu_0$ was evaluated from this distribution using a one-sided test.
%
The right panel of Fig.~\ref{fig:Pin_Pdp_plocal} shows the resulting $\PDP$ and local $p$-values.
The minimum local $p$-value was $0.33$.
Thus, no significant excess was found in the search band.
%
%
%
%

The systematic uncertainties were evaluated following the framework established in the previous DOSUE-RR searches~\cite{DOSUE-K,DOSUE-J}, with modifications for this work.
Table~\ref{tab:syst} lists the contributions to the systematic uncertainty in $\chi$.
The uncertainty in $\Aeff$ was estimated from the discrepancy between the measured and simulated HPBWs, as described above.
The uncertainty in $\chi$ from gain includes contributions from the uncertainty in the blackbody source temperature (0.69\%),
the imperfect emissivity of the blackbody sources (0.42\%),
the residual frequency deviation of the gain from its linear fit (0.62\%),
and time variation of the gain, estimated from the difference between the calibrations before and after the measurement (2.7\%).
LSB leakage during gain calibration leads to a systematic overestimate of the gain for a DPDM signal in the USB.
Given the IRR of $39.35\dB$ at $242.95\GHz$, the resulting systematic uncertainty in $\chi$ was negligible ($<0.1\%$).
Frequency binning introduces a fitting bias that depends on the position of the signal peak within a bin.
This bias was estimated from the variation in $\PDP$ when the signal frequency was shifted by $\pm\dnu/2 = \pm10\kHz$ in a simulation.
The uncertainties from the antenna pointing direction relative to the metal plate and from the direction of the conversion photons were estimated using the measured alignment and the antenna beam width.
The measured tilt of the metal plate relative to the antenna was at most $0.72\Deg$.
Owing to the relatively broad antenna beam, this misalignment contributes only a small systematic uncertainty in $\chi$.
As noted in Sect.~1, the conversion photons propagate within $\lesssim 0.1\Deg$ of the surface normal.
For the same reason, this deviation contributes a negligible systematic uncertainty.
The speed of the Earth in the Galactic frame on the observation date was $|\mathbf{v}_{\rm E}|=219.61\kmpers$~\cite{DMVelocity}.
The systematic uncertainty in $\chi$ due to the difference from the nominal value of $220\kmpers$ used in the signal model was negligible.
For the dark matter density, we used the uncertainty reported in Ref.~\cite{DMdensity}.

\begin{table}[t]
\centering
\caption{\label{tab:syst}
Systematic uncertainties in the coupling constant $\chi$.}
\begin{tabular}{lc}
\hline\hline
Source & Uncertainty (\%) \\
\hline
Effective aperture area ($\Aeff$) & 2.4 \\
Gain & 2.9 \\
LSB leakage & $<0.1$ \\
Frequency binning & 2.2 \\
Antenna pointing direction & 0.35 \\
Direction of conversion photons & $<0.1$\\
Earth speed ($\vE$) & $<0.1$ \\
Dark matter density ($\rho$) & 3.8 \\
\hline
Total & 5.8 \\
\hline\hline
\end{tabular}
\end{table}

The upper bound on $\PDP$ at the 95\% confidence level was calculated as
\begin{equation}
 \PDP^{95}=\max(0,\PDP)+1.89 \sigma ,
 \label{eq:pdp_limit}
\end{equation}
where the coefficient 1.89 corresponds to the one-sided 95\% quantile of the null-sample distribution of $x$.
The upper limits on $\PDP$ were converted into limits on $\chi$ using Eq.~(\ref{eq:power}).
The total systematic uncertainty listed in Table~\ref{tab:syst} was also incorporated into the upper limits on $\chi$.

\begin{figure}[t]
\centering
\includegraphics[width=0.9\columnwidth]{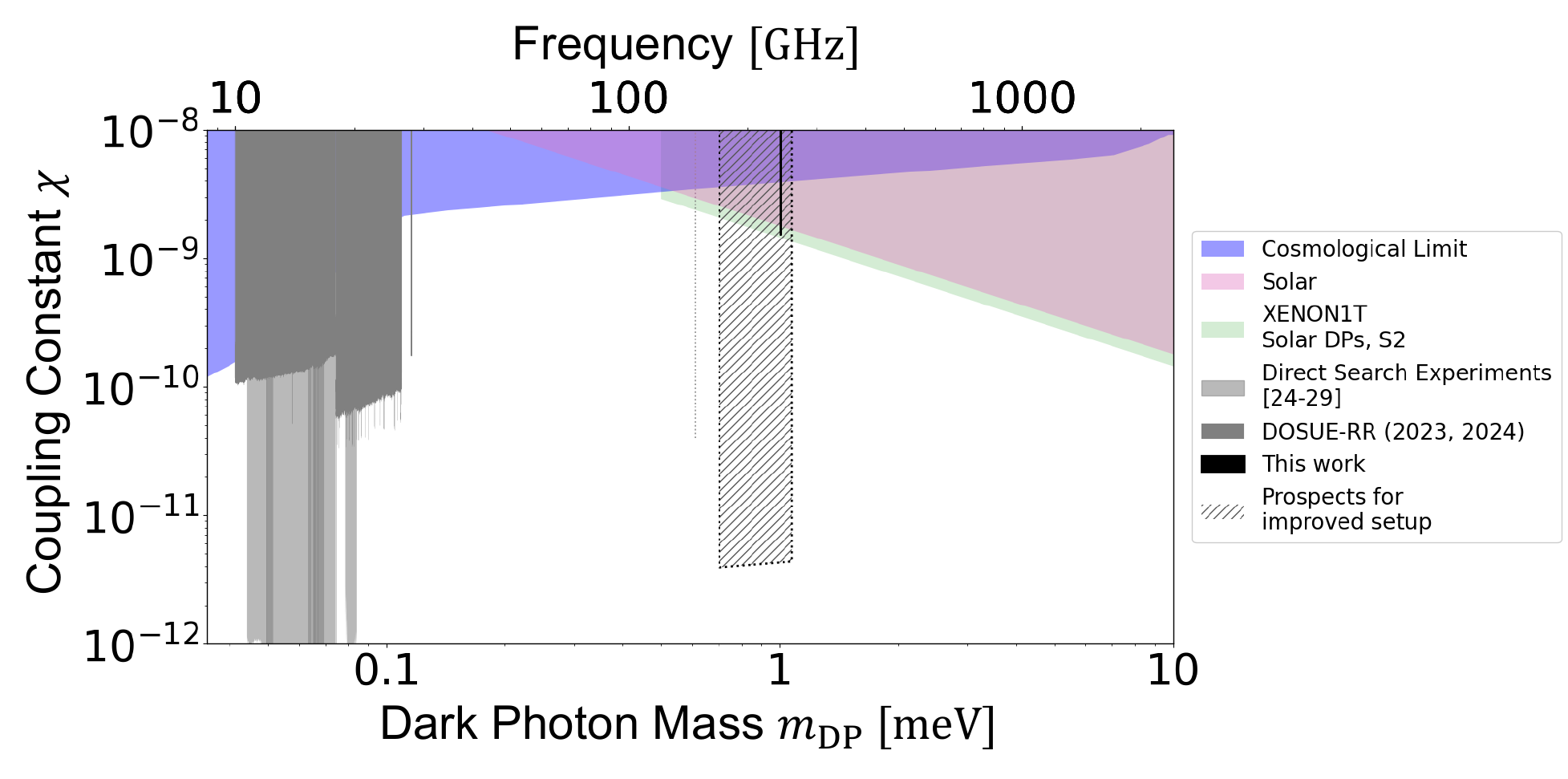}
\par\medskip
\includegraphics[width=0.75\columnwidth]{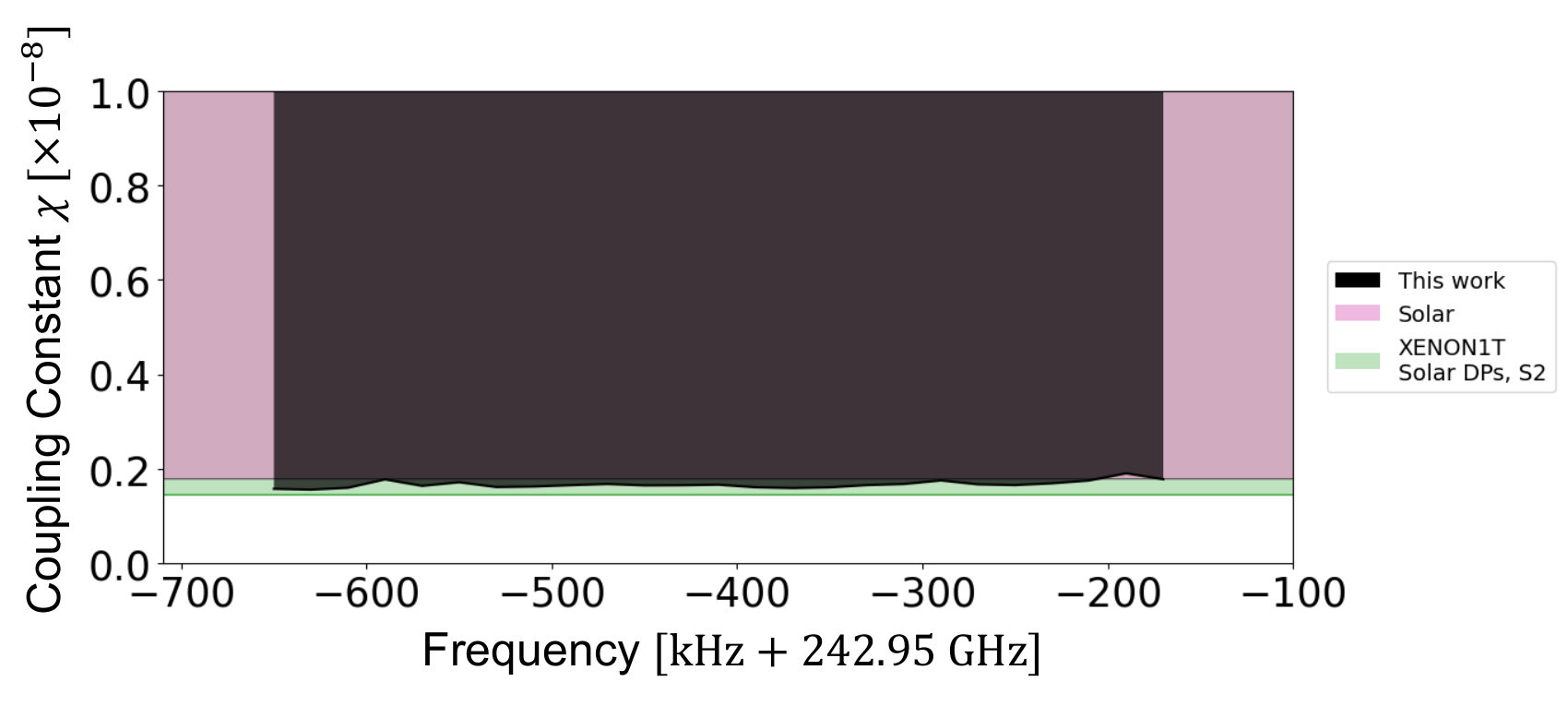}
\caption{\label{fig:limit}
Constraints on $\chi$ as a function of $\mDP$~\cite{Tomita,ORPHEUS,QuantumCyclotron,DOSUE-K,BRASSp,GigaBREAD,DOSUE-J,MADMAX,solar,Wispy,XENON1T-S2,AxionLimit, AxionLimitsZenodo}.
The black line shows the upper limits from the present search at the 95\% confidence level near $242.95\GHz$.
The hatched region shows the projected sensitivity of the upgraded setup over \ymassband{} (corresponding to \yband).
The lower panel is an enlarged view of the region around the present search.
}
\end{figure}

\textit{5. Results.  }
Figure~\ref{fig:limit} shows the resulting upper limits together with existing constraints.
The present data set yields upper limits of $\chi < \chilimit$ at the 95\% confidence level in the mass range \massband.
With only one hour of data taking, the present limits are comparable to existing constraints from solar observations and XENON1T in this mass range~\cite{solar,XENON1T-S2}.
This shows that an SIS-mixer receiver is a viable platform for direct DPDM searches in the $\meV$ mass range.

\textit{6. Future prospects.  }
Future searches will improve the sensitivity and extend the mass coverage.

To improve the sensitivity, we will introduce a parabolic focusing mirror to collect more conversion photons.
In the present setup, the collected power is limited by the aperture of the horn antenna (diameter: $7.1 \mm$, $\Aeff = 18.4 \sqmm$).
The mirror is currently under construction.
With this mirror, the setup is expected to have an effective aperture area equivalent to that of a circle approximately $300\mm$ in diameter.
This corresponds to an increase in $\Aeff$ by a factor of $\pi (150\mm)^2/18.4\sqmm \simeq 3800$.
For the configuration with the focusing mirror, we assume $\Trx=140\K$ and an additional thermal loading temperature of $40\K$.
This receiver noise temperature is the same as that achieved in the present measurement.
The loading temperature is expected to be much lower than room temperature because, with the focusing mirror, the antenna beam couples predominantly to the low-emissivity metal plate, which reflects it back to the 4 K stage or the 40 K chamber.

To extend the frequency coverage, we have developed a broadband spectrometer~\cite{dSpec}.
While the present setup has an instantaneous bandwidth of only $2.5\MHz$, the new spectrometer provides an instantaneous bandwidth of approximately $4\GHz$.
This upgrade will allow us to scan a wide frequency range efficiently.
In addition, we will develop 2SB filters with different separation frequencies to cover other RF bands.
We also plan to measure the USB and LSB simultaneously, whereas only the USB was used in the present measurement.

By combining these upgrades and assuming the receiver and loading temperatures given above, we expect to achieve a sensitivity of $\chi \lesssim 3.9\times10^{-12}$ over a broad DPDM mass range (\ymassband, corresponding to \yband) with a one-day measurement for each $8\GHz$ band using two spectrometers, as shown in Fig.~\ref{fig:limit}.

\textit{7. Conclusion.  }
We have demonstrated a direct search for DPDM using a cryogenic millimeter-wave receiver based on an SIS mixer.
We searched the mass range \massband, corresponding to the frequency range \searchband.
No significant signal was found, and we set upper limits of $\chi < \chilimit$ at the 95\% confidence level.
These results demonstrate that SIS-mixer receivers are a promising platform for direct DPDM searches in the $\meV$ mass range.
Future upgrades with a focusing mirror and a broadband spectrometer will extend the search to a wider mass range with higher sensitivity.

\ack
This work was supported
by JSPS KAKENHI under Grant Numbers 23H00111, 21H05460, and 22K18713;
by JST CREST, Japan, Grant Number JPMJCR23I3;
and by JST SPRING, Grant Number JPMJSP2110.
S.A. acknowledges the Hakubi Project of Kyoto University.
H.T. is supported by a JSPS Research Fellowship for Young Scientists and JSPS KAKENHI Grant Number JP24KJ1452.

\bibliography{references}

\begin{thebibliography}{10}

\bibitem{Wispy}
P.~Arias et~al., J. Cosmol. Astropart. Phys., {\bf 2012}(06), 013 (2012).

\bibitem{inflation}
P.~W. Graham, J.~Mardon, and S.~Rajendran, Phys. Rev. D, {\bf 93}, 103520 (2016).

\bibitem{AxionLimit}
A.~Caputo, A.~J. Millar, C.~A.~J. O'Hare, and E.~Vitagliano, Phys. Rev. D, {\bf 104}, 095029 (2021).

\bibitem{DishAntenna}
D.~Horns et~al., J. Cosmol. Astropart. Phys., {\bf 2013}(04), 016 (2013).

\bibitem{DishDirection}
J.~Jaeckel and S.~Knirck, J. Cosmol. Astropart. Phys., {\bf 2016}(01), 005 (2016).

\bibitem{PhysRevLett.51.1415}
P.~Sikivie, Phys. Rev. Lett., {\bf 51}, 1415--1417 (1983).

\bibitem{DMdensity}
R.~Catena and P.~Ullio, J. Cosmol. Astropart. Phys., {\bf 2010}(08), 004 (2010).

\bibitem{solar}
N.~Vinyoles et~al., J. Cosmol. Astropart. Phys., {\bf 2015}(10), 015 (2015).

\bibitem{PlaneParabolic}
S.~Knirck et~al., J. Cosmol. Astropart. Phys., {\bf 2018}(11), 031 (2018).

\bibitem{Dandelion}
I.~Ourahou et~al.,
\newblock {First Dark Photon Search Results from the Dandelion Experiment} (2026),  {{arXiv:2602.18218}}.

\bibitem{DOSUE-K}
S.~Kotaka et~al. (DOSUE-RR Collaboration), Phys. Rev. Lett., {\bf 130}, 071805 (2023).

\bibitem{VDI_LNA}
{Virginia Diodes, Inc.},
\newblock Low noise amplifiers: Operational manual,
\newblock \url{https://vadiodes.com/wp-content/uploads/2026/02/VDI-795.38-Low-Noise-Amplifier-Product-Manual.pdf} (2026),
\newblock Revision February 25, 2026, p. 4. Accessed October 7, 2026.

\bibitem{TuckerFeldman}
J.~R. Tucker and M.~J. Feldman, Rev. Mod. Phys., {\bf 57}, 1055--1113 (1985).

\bibitem{SIS2007}
T.~Nakajima et~al., Publ. Astron. Soc. Japan, {\bf 59}(5), 1005--1016 (2007).

\bibitem{2SBF}
Y.~Hasegawa et~al., Publ. Astron. Soc. Japan, {\bf 69}(6), 91 (2017).

\bibitem{Asayama2015IRF}
S.~Asayama, Y.~Hasegawa, A.~Mizuno, H.~Ogawa, and T.~Onishi, J. Infrared Millim. Terahertz Waves, {\bf 36}(5), 445--454 (2015).

\bibitem{KeysightE8257D}
{Keysight Technologies},
\newblock {E8257D PSG Microwave Analog Signal Generator: Data Sheet},
\newblock \url{https://www.keysight.com/content/dam/keysight/en/doc/ungate/data-sheets/5989-0698.pdf} (2025),
\newblock Publication no. 5989-0698EN, Accessed August 21, 2026.

\bibitem{RTMLI}
J.~Choi et~al., Rev. Sci. Instrum., {\bf 84}, 114502 (2013).

\bibitem{DOSUE-J}
S.~Adachi et~al. (DOSUE-RR Collaboration), Phys. Rev. D, {\bf 109}, 012008 (2024).

\bibitem{DMVelocityDistribution}
A.~K. Drukier, K.~Freese, and D.~N. Spergel, Phys. Rev. D, {\bf 33}, 3495--3508 (1986).

\bibitem{DMVelocity}
F.~Mayet et~al., Phys. Rep., {\bf 627}, 1--49 (2016).

\bibitem{XENON1T-S2}
H.~An, M.~Pospelov, J.~Pradler, and A.~Ritz, Phys. Rev. D, {\bf 102}, 115022 (2020).

\bibitem{dSpec}
H.~Takeuchi, J.~Suzuki, S.~Adachi, and O.~Tajima, Prog. Theor. Exp. Phys., {\bf 2025}(5), 053H02 (2025).

\bibitem{Tomita}
N.~Tomita et~al., J. Cosmol. Astropart. Phys., {\bf 2020}(09), 012 (2020).

\bibitem{ORPHEUS}
R.~Cervantes et~al., Phys. Rev. D, {\bf 106}, 102002 (2022).

\bibitem{QuantumCyclotron}
X.~Fan et~al., Phys. Rev. Lett., {\bf 129}, 261801 (2022).

\bibitem{BRASSp}
F.~Bajjali et~al., J. Cosmol. Astropart. Phys., {\bf 2023}(08), 077 (2023).

\bibitem{GigaBREAD}
S.~Knirck et~al. (BREAD Collaboration), Phys. Rev. Lett., {\bf 132}, 131004 (2024).

\bibitem{MADMAX}
J.~Egge et~al. (MADMAX Collaboration), Phys. Rev. Lett., {\bf 134}, 151004 (2025).

\bibitem{AxionLimitsZenodo}
C.~O'Hare,
\newblock {cajohare/AxionLimits: AxionLimits},
\newblock \url{https://cajohare.github.io/AxionLimits/} (July 2020).

\end{thebibliography}

\end{document}